# Cellular liberality is measurable as Lempel-Ziv complexity of fastq files

Norichika OGATA
*Nihon BioData Corp.*
Kawasaki, Japan
*Tokyo University of Agriculture and Technology*
Tokyo, Japan
*Manufacturing Technology Association of Biologics*
Kobe, Japan
norichik@nbiodata.com
https://orcid.org/0000-0002-6856-6655

Aoi HOSAKA
*Nihon BioData Corp.*
Kawasaki, Japan
*Kihara Institute for Biological Research*
Yokohama, Japan
hosak@nbiodata.com
https://orcid.org/0000-0003-0494-3359

*Abstract*—Many studies used the Shannon entropy of transcriptome data to determine cell dedifferentiation and differentiation. The collection of evidence has strengthened the certainty that the transcriptome's Shannon entropy may be used to quantify cellular dedifferentiation and differentiation. Quantifying this cellular status is being justified, we propose the term "liberality" for the quantitative value of cellular dedifferentiation and differentiation. In previous studies, we must convert the raw transcriptome data into quantitative transcriptome data through mapping, tag counting, assembling, and more bioinformatic processing to calculate the liberality. If we could remove this conversion step from estimating liberality, we could save computing resources and time and remove technical difficulties in using the computer. In this study, we propose a method of calculating cellular liberality without those transcriptome data conversion processes. We could calculate liberality by measuring the compression rate of raw transcriptome data. This technique, independent of reference genome data, increased the generality of cellular liberality.

*Keywords—Lempel-Ziv complexity, Shannon entropy, liberality, dedifferentiation, differentiation, transcriptome, RNA-seq, genomics, compression, classification, clustering*

## I. Introduction

Cellular dedifferentiation and differentiation have been understood as the direction of cellular morphology and phenotype change [1, 2]. In this decade, several studies [3-7] following our research [8] repeatedly measured the degree of cellular dedifferentiation and differentiation as a transcriptome Shannon entropy. The Shannon entropy is a kind of alpha diversity in ecology [9], and the transcriptome Shannon entropy is simply transcriptome diversity [10, 11]. It is not incorrect to call it the alpha diversity of the transcriptome, but that would leave its biological significance undefined, as would each principal component that came up in the principal component analysis. Since we can quantitatively assess, judge, and define that dedifferentiation is an increase in the Shannon entropy of the transcriptome and differentiation is a decrease in the Shannon entropy of the transcriptome, it is more accurate to position the "value of information entropy of the transcriptome" not as a mere bioinformatics measure; however, as a number with obvious biological and bioengineering significance, such as viable cell rate, cell density, specific growth rate, or pcd (pg/cell/day). Here we call the quantitative value of cellular dedifferentiation and differentiation "liberality," since a previous study explained the changes were happening to cultured cells as "libère" [12].

The term transcriptome data is a little confusing, which is understood as the entire RNA in a biological sample; cells, tissues, organisms or environments. The transcriptome is primarily derived as text data of nucleotide sequences, which is sequenced using the next generation sequencer. Those sequence data are usually converted into quantitative data through a bioinformatic process called mapping or alignment. The textual data of nucleotide sequences and the quantified amount of each RNA is referred to as transcriptome data.

We want to calculate the liberality without the transcriptome quantifying process. Incorporating the transcriptome quantification procedure into the liberality calculation process not only consumes computer resources and is cumbersome to deal with but also reduces the generality of the cellular liberality. The transcriptome quantification process essentially depends on referential genome sequence data. The referential genome sequence data is still imperfect [13, 14]. Genomic data are being improved as genomic technology develops; for example, even the human genome, where the most effort is being made, is amid revision [15]. In a practical sense, the referential genome sequence data cannot be uniquely defined. By removing the referential genome sequence data from the liberality measuring process, we attempted to increase the generality of the liberality.

The Shannon entropy is similar to the Lempel-Ziv complexity [16]. In a previous study, we demonstrated the Lempel-Ziv complexity of transcriptome data calculated by measuring the compression rate of the files is helpful like the Shannon entropy [17]. In that study, we used quantified transcriptome data. We cannot measure the Shannon entropy of the raw transcriptome data but there is a possibility that we can measure the Lempel-Ziv complexity of the raw transcriptome data. We can estimate Lempel-Ziv complexity by compressing given individual files. Therefore, we can possibly measure the cellular liberality by simply measuring the compression rate of the text file. Previous studies calculated Lempel-Ziv complexity of biological data for classifying [18-20]. Those studies did not position Lempel-Ziv complexity of of biological data as a number with any obvious biological significance. It was only a mere bioinformatics measure for classifying.

In this study, we demonstrate a liberality method by measuring the compression rate of raw textual transcriptome data using wheat leaf transcriptome data. In the first experiment, we compared liberalities measured in two different processes; one process included the transcriptome quantification processes. The Shannon entropy of the quantified transcriptome data was determined. The compression rate of raw textual transcriptome data was investigated in the other phase. The liberties measured by the various methods were consistent. To test the method's robustness, we picked 10k, 1k, and 100 sequence read from raw sequence data and estimated the compression rate. Additionally, we assessed the equivalency between the Shannon entropy of quantified transcriptome data and the Lempel-Ziv complexity of raw textual transcriptome data using Chinese hamster ovary (CHO) cells and silkworm transcriptome data. Our advanced method worked more when applied to those data.

## II. MATERIAL AND METHODS

### A. Dataset Description

We used publicly available transcriptome sequencing datasets from the DDBJ's public sequence data repository (Sequence Read Archive, https://www.ddbj.nig.ac.jp/dra/index.html). We used three datasets, wheat leaf transcriptomes (DRA008774) [21], Chinese hamster ovary (CHO) cells transcriptomes (DRA006016) [22] and silkworm fat body transcriptomes (DRA002853) [11].

### B. Dataset Description (Wheat Leaf)

The wheat leaf transcriptomes were sequenced and achieved by other scientists. To track changes in gene expression during leaf development, they cut wheat leaves into 18 samples from leaf base to leaf tip, 14 of which were subjected to sequencing. The distance of cells from the leaf base is related to developmental time in wheat leaf development. Cells are anticipated to change their liberality during development; we thought that changes in liberality could be easily monitored by employing a dataset that follows this developmental process. DRR187484, DRR187485, DRR187486, DRR187487, DRR187488, DRR187489, DRR187490, DRR187491, DRR187492, DRR187493, DRR187494, DRR187496, DRR187497, DRR187498, DRR187499, DRR187500, DRR187501, DRR187502, DRR187503, DRR187504, DRR187505, DRR187506, DRR187507, DRR187508, DRR187509, DRR187511, DRR187512, DRR187513, DRR187514, DRR187515, DRR187516, DRR187517, DRR187518, DRR187519, DRR187520, DRR187521, DRR187522, DRR187523, DRR187524, DRR187526, DRR187527, and DRR187528 were used. Sequencing conditions were unknown.

### C. Dataset Description (CHO cells)

The CHO cell transcriptomes were sequenced in our laboratory. We cultured CHO cells in flasks and obtained cells at 3, 5, 6, and 7 days in culture. In this experiment, we have already noticed that the liberality of the cells decreases throughout the culture. Because it is commercially advantageous to understand the liberality for manufacturing control in cell culture for industry, it is critical to demonstrate that the method of measuring liberality proposed in this study applies to the culture of CHO cells, the most common cells in biopharmaceutical manufacturing [23]. DRR099453, DRR099454, DRR099455, DRR099465, DRR099466, DRR099467, DRR099477, DRR099478, DRR099479, DRR099489, DRR099490, and DRR099491 were used. We sequenced them in a single run.

### D. Dataset Description (Silkworm Fat Bodies)

Our lab also sequenced the transcriptomes of silkworm fat bodies. We collected silkworm fat bodies to confirm if cell liberality demonstrates hysteresis in the response of cells to environmental changes. We assessed differences between cells before being given the drug and cells that were given the medication and then had the drug taken away. In the simple-dose samples, silkworm fat bodies were cultured for 80 h in phenobarbital–non-supplemented MGM-450 insect medium followed by 10 h in MGM-450 insect medium supplemented with 0, 0.25, 1.0, 2.5, and 12.5 mM phenobarbital after cultivation. In the hysteresis samples, silkworm fat bodies were cultured for 10 h in MGM-450 insect medium supplemented with 0 and 0.25 mM phenobarbital after 90 h previous cultivation (80 h in phenobarbital–non-supplemented MGM-450 insect medium followed by 10 h in 1.0 mM phenobarbital-supplemented MGM-450 insect medium). Understanding the liberality of cells is also valuable for managing studies in which cells are drugged and their responses are examined [24]. We would like to indicate that the proposed method can also be used for this purpose. DRR027746, DRR027747, DRR027748, DRR027749, DRR027750, DRR027751, DRR027752, DRR027753, DRR027754, DRR027755, DRR027756, DRR027757, DRR027758, DRR027759, DRR027760, DRR027761, DRR027762, DRR027763, DRR027764, DRR027765, DRR027766, DRR027767, DRR027768, and DRR027769 were used. We processed their nucleotides in different DNA workflows and sequenced in several runs.

### E. Liberality Calculation as the Shannon Entropy of Quantified Transcriptome Data

As previously explained, we quantified each transcriptome data by mapping, assembling and tag counting [8, 11, 22, 23, 24]. The Shannon entropy of each quantified transcriptome data was calculated as previously described [8, 11].

All short reads of the wheat leaf transcriptomes were mapped to the wheat genome (iwgsc_refseqv2.1_assembly.fa and iwgsc_refseqv2.1_annotation_200916_HC.gtf) using STAR. Read counts were processed using featureCounts.

The quality of the raw reads of the CHO transcriptomes was analyzed with FastQC (version 0.11.3). All short reads were mapped to the CHO-K1 RefSeq assembly (22,516 sequences; RefSeq Assembly ID: GCF_000223135.1) and CHO-K1 mitochondrial DNA (1 sequence; RefSeq Assembly ID: GCF_000055695.1) using Bowtie2 (version 2.3.4.1) and quantified using RSEM (version 1.2.31).

The silkworm fat body's sequence read qualities were controlled using the FastQC program. Short-read sequences were mapped to an annotated silkworm transcript sequence obtained from KAIKObase (http://sgp.dna.affrc.go.jp) using the Bowtie program. A maximum of two mapping errors were allowed for each alignment.

TABLE I. ANALYSES PERFORMED IN THIS STUDY

| Dataset | Calculated Liberality | | | |
| --- | --- | --- | --- | --- |
| | Shannon Entropy | Lempel-Ziv Complexity (ID and QV removed fastq files) | Lempel-Ziv Complexity (native fastq files) | Lempel-Ziv Complexity (ID and QV removed small fastq files) |
| Wheat leaf | Fig. 1a, Fig. 1d | Fig. 1b, Fig. 1d | Fig. 1c | Fig. 2 |
| CHO cells | Fig. 3a | Fig. 3b | Fig. 3c | - |
| Silkworm fat bodies | Fig. 4a | Fig. 4b | Fig. 4c | - |

*F. Liberality Calculation as the Lempel-Ziv Complexity of Raw Transcriptome Data*

We measured the compression rate of each sequence data by compassion between file sizes of each uncompressed and compressed sequence textual data. The compression was performed using gzip (Wheat leaf), zip (CHO, ID, and QV removed fastq), bzip2 (CHO, native fastq) and compress (silkworm) command in the UNIX. The file sizes measurement was performed using a command with "-al" options in the UNIX. We removed ID lines and quality value (QV) lines from fastq data using the awk command with "(NR%4==2){print}". We processed ID and QV and removed fastq files and native fastq files.

*G. Robustness Testing for Sequence Reads the Amount*

The fastq data derived recently generally contains more than 10,000,000 nucleotide sequence reads. We confirmed that the method proposed in this study works well even when a small number of sequences are extracted from these data sets and confirmed the method's robustness. We randomly sampled 100, 1000, and 10000 reads from the wheat leaf transcriptomes and measured liberality as the Lempel-Ziv complexity of raw transcriptome data. We processed ID and QV removed small fastq files.

## III. RESULTS AND DISCUSSION

*A. The Wheat Leaf Transcriptome*

The liberalities calculated as the Shannon entropy of quantified transcriptome data (old method) and those calculated as the Lempel-Ziv complexity of raw transcriptome data (ID and QV removed fastq files) using the method proposed in this study were consistent, both in correspondence with other biological parameters (developmental time, Figs. 1a, 1b, 1c) and in direct comparison with each other (Fig. 1d). The Lempel-Ziv complexity of raw transcriptome data (native fastq files) showed several faulty values (Fig. 1c). The QV in the fastq files depend on cluster density in the flow cells in the sequencing run. The ID in the fastq files is also determined by the sequencing run. This is speculation; samples processed in the same lot of DNA workflow would be analyzed using liberalities directly estimated from native fastq files.

The liberalities calculated as the Lempel-Ziv complexity of small raw transcriptome data (ID and QV removed small fastq files) and that of full raw transcriptome data (ID and QV removed fastq files) were partially consistent in direct comparison with each other. The method proposed in this study was robust; we could calculate liberalities from only

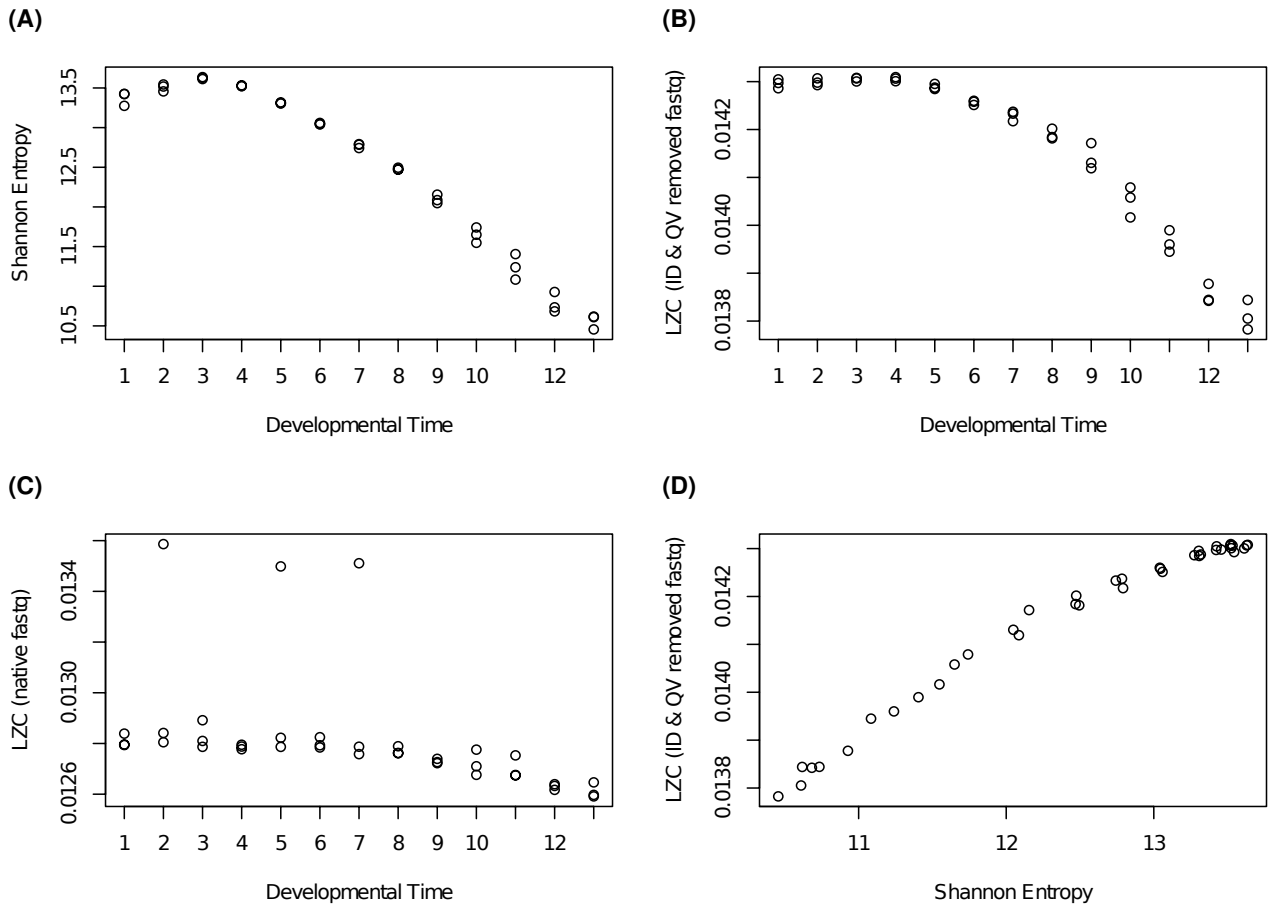

Fig. 1. Comparison between liberalities of wheat leaf calculated in different methods.
The liberalities of the wheat leaf transcriptome are calculated as the Shannon entropy of quantified transcriptome data and that as the Lempel-Ziv complexity of raw transcriptome data. (a) Scatter plot of developmental time vs. Shannon entropy of quantified transcriptome data. (b) Scatter plot of developmental time vs. Lempel-Ziv complexity of raw transcriptome data (ID and QV removed fastq files). (c) Scatter plot of developmental time vs. Lempel-Ziv complexity of raw transcriptome data (native fastq files). (d) Scatter plot of Shannon entropy of quantified transcriptome data vs. Lempel-Ziv complexity of raw transcriptome data (ID and QV removed fastq files).

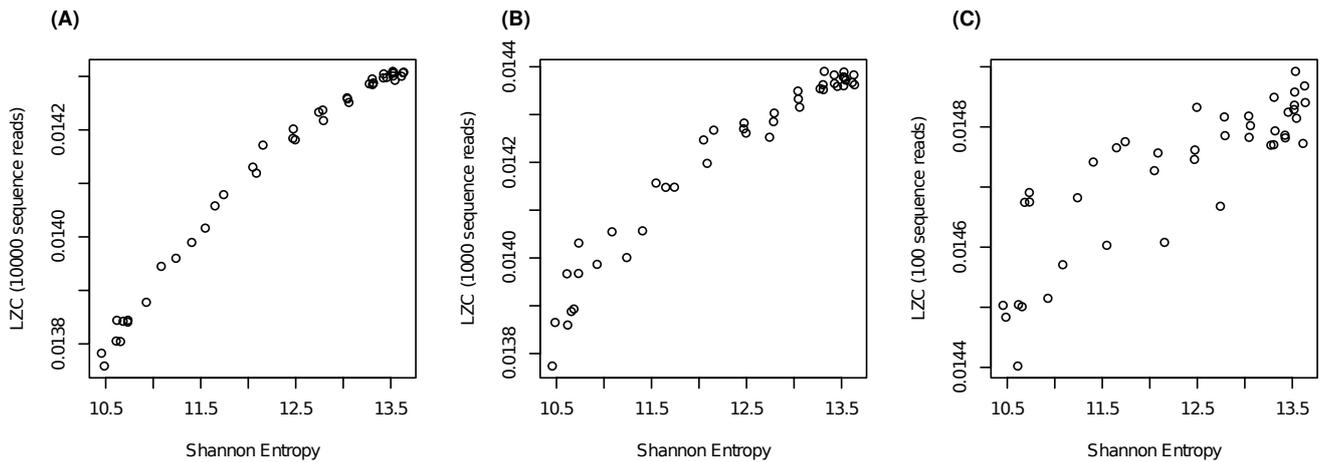

Fig. 2. Comparison between liberalities calculated in different data amounts.

The liberalities calculated as the Shannon entropy of quantified transcriptome data (old method) and that calculated as the Lempel-Ziv complexity of small raw transcriptome data (10000, 1000, and 100 sequence reads, ID and QV removed fastq files) using the method proposed in this study were compared directly. (a) Scatter plot of Shannon entropy vs. Lempel-Ziv complexity of small raw transcriptome data (10000 reads). (b) Scatter plot of Shannon entropy vs. Lempel-Ziv complexity of small raw transcriptome data (1000 reads). (c) Scatter plot of Shannon entropy vs. Lempel-Ziv complexity of very small raw transcriptome data (100 reads).

1000 sequence reads (This is under 0.01 % of generally obtained transcriptome sequence data.). One hundred sequence reads were not enough to estimate liberalities.

### B. The CHO Cells Transcriptome

The liberalities calculated in various approaches were consistent, both in correspondence with other biological parameters (culture time, Figs. 3a, 3b, 3c) and in direct comparison with each other ($r^2 = 0.92$, SE and KC (ID and QV removed fastq), $r^2 = 0.93$, SE, and KC (native fastq)). The Lempel-Ziv complexity of raw transcriptome data (native fastq files) did not indicate faulty values appearing in the wheat leaf transcriptome (Figs. 1c and 2c). These samples are processed in the same RNA/DNA workflow and in the same sequencing run and could be analyzed using liberalities directly calculated from native fastq files.

### C. The Silkworm Fat Bodies Transcriptome

The liberalities calculated in several methods partially agreed with the correspondence with other biological parameters (drug concentration, Figs. 4a and 4b). The liberality was determined by the drug concentration at that time and previous drug concentrations. This is a hysteretic phenomenon, and a hysteretic phenomenon provides evidence of a bi-stable system. These results indicate the multi-stability of the genome expression system. Notably, the liberalities calculated as the Lempel-Ziv complexity of raw transcriptome data (ID and QV removed fastq) were categorized into two parts clear (Fig. 4b). The Lempel-Ziv complexity of raw

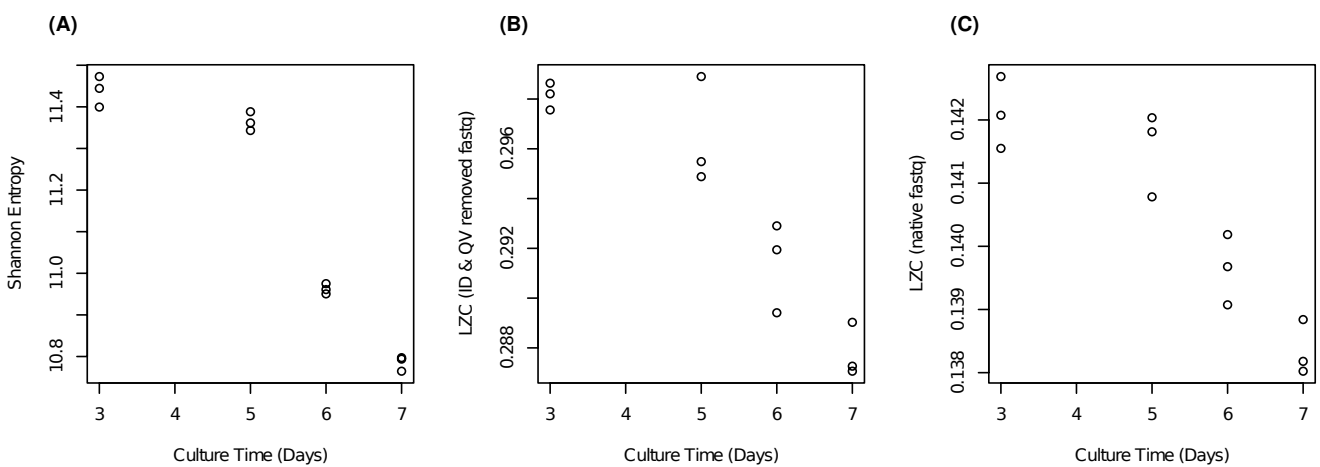

Fig. 3. Comparison between liberalities of CHO cells calculated in various methods.

The liberalities of the Chinese hamster ovary cells are calculated as the Shannon entropy of quantified transcriptome data and that as the Lempel-Ziv complexity of raw transcriptome data. (a) Scatter plot of culture time (days) vs. Shannon entropy of quantified transcriptome data. (b) Scatter plot of culture time vs. Lempel-Ziv complexity of raw transcriptome data (ID AND QV removed fastq files). (c) Scatter plot of culture time vs. Lempel-Ziv complexity of raw transcriptome data (native fastq files).

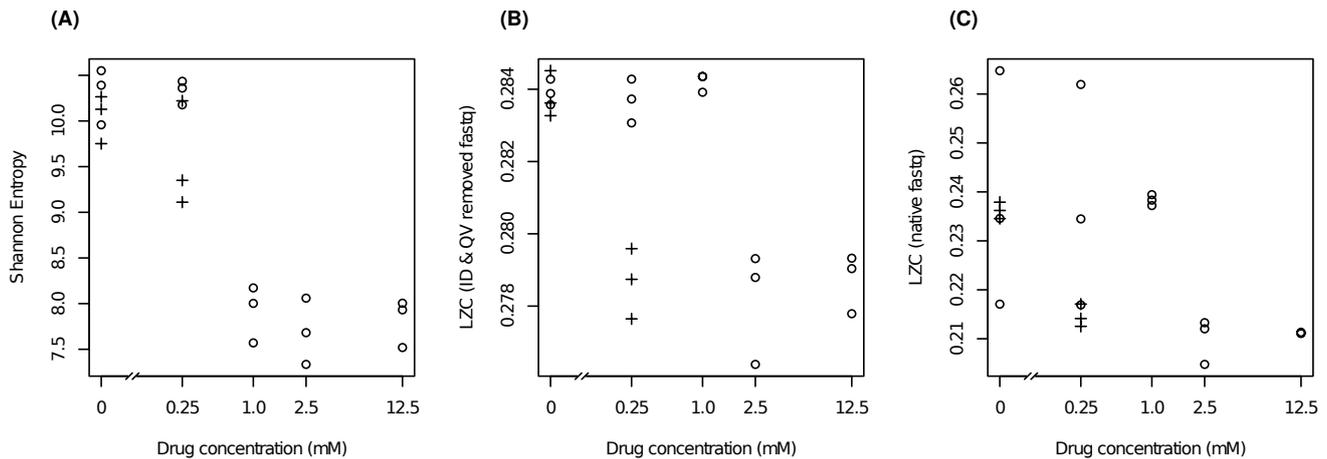

Fig. 4. Comparison between liberalities of silkworm fat bodies calculated in different methods.

The liberalities of the silkworm fat body are calculated as the Shannon entropy of quantified transcriptome data and that as the Lempel-Ziv complexity of raw transcriptome data. (a) Scatter plot of drug concentration vs. Shannon entropy of quantified transcriptome data. (b) Scatter plot of drug concentration vs. Lempel-Ziv complexity of raw transcriptome data (ID and QV removed fastq files). (c) Scatter plot of drug concentration vs. Lempel-Ziv complexity of raw transcriptome data (native fastq files).

transcriptome data (native fastq files) were different from the Shannon entropy and the Lempel-Ziv complexity of raw transcriptome data (ID and QV removed fastq files) (Fig. 4). These samples are processed in the various DNA workflow and sequencing run, and could not be evaluated using liberalities directly estimated from native fastq files.

## IV. CONCLUSION

We had the perception that transcriptome sequence data file compression ratios were significant a dozen years ago, but we did not receive a clear indication. This study shows that the Lempel-Ziv complexity obtained from the file compression ratio is helpful in estimating cellular liberality using three data sets. We could calculate liberalities without the genome data. This increased the generality of the cellular liberality. In this experiment, the Shannon entropy and the Lempel-Ziv complexity were consistent in the wheat leaf transcriptome and CHO cells transcriptome but were not in the silkworm transcriptome. Since only the silkworm data are old, it is believed that recent developments in transcriptome measurement technology have reduced the noise in the transcriptome data. Advancements in DNA/RNA sample prep protocols have improved reproducibility, and advancements in sequencing equipment have reduced sequencing errors and QV variation. These technological developments have probably expanded the effective range of classification by compression [25]. As genomics measuring technology advances, the ability to extract signals and messages from informatics analysis grows, and informatics analysis will become increasingly significant in the future.

## ACKNOWLEDGMENT

We are grateful Prof. bot22610669 for his discussions on twitter. He told me that it is common knowledge among compression experts that compression and classification (as well as similarity quantification) are two sides of the same coin [26].

## AUTHOR CONTRIBUTION

AJH analyzed the wheat leaf transcriptome data. NO did all other things.